\documentclass[12pt,preprint]{aastex}

\shorttitle{The eclipsing binary V209 in $\omega$~Cen
}
\shortauthors{Kaluzny et al.}

\begin{document}

\title{The Clusters AgeS Experiment (CASE): \\ 
V209 $\omega$~Cen -- An Eclipsing Post-Common Envelope Binary \\ in 
the Globular Cluster $\omega$~Cen\footnote{This paper utilizes data 
obtained with the 6.5-meter Magellan Telescopes located at Las Campanas 
Observatory, Chile.}
}

\author{J.~Kaluzny\altaffilmark{2}, S.~M.~Rucinski\altaffilmark{3},
I.~B.~Thompson\altaffilmark{4}, W.~Pych\altaffilmark{2}
and W.~Krzeminski\altaffilmark{5}
}

\altaffiltext{2}{Copernicus Astronomical Center, Bartycka 18,
00-716 Warsaw, Poland; (jka,pych)@camk.edu.pl}

\altaffiltext{3}{David Dunlap Observatory, Department of Astronomy and 
Astrophysics, University of Toronto, P.O. Box 360, Richmond Hill, 
ON L4C 4Y6, Canada; rucinski@astro.utoronto.ca}

\altaffiltext{4}{Carnegie Observatories, 813 Santa Barbara St.,
Pasadena, CA 91101-1292; ian@ociw.edu}

\altaffiltext{5}{Las Campanas Observatory, Casilla 601, La Serena, Chile; 
wojtek@lco.cl}

\begin{abstract}
We use photometric and spectroscopic observations of the detached eclipsing
binary V209~$\omega$~Cen to derive the masses, radii, and luminosities
of the component stars. The system exhibits total eclipses and,
based on the measured systemic velocity and the derived distance,
is a member of the globular cluster $\omega$~Cen.
We obtain $0.945\pm 0.043\,M_{\odot}$, 
$0.983\pm 0.015\,R_{\odot}$ and $6.68\pm 0.88\,L_{\odot}$
for the cooler, but larger and more luminous primary component.  
The secondary component has $0.144\pm 0.008\,M_{\odot}$,
$0.425\pm 0.008\,R_{\odot}$ and $2.26\pm 0.28\,L_{\odot}$. The
effective temperatures are estimated at 9370~K for the
primary and at 10866~K for the secondary. 
On the color-magnitude diagram of the cluster, 
the primary component occupies a position between the tip 
of the blue straggler region and the extended horizontal branch
while the secondary component is located close to the red
border of the area occupied by hot subdwarfs. However, its radius 
is too large and its effective temperature is too low for
it to be an sdB star.
We propose a scenario leading to the formation of a system with such 
unusual properties with the primary component ``re-born''
from a former white dwarf which accreted a new envelope
through  mass transfer from its companion. The secondary star
has lost most of its envelope while starting its ascent onto 
the sub-giant branch. It failed to ignite helium in its core and is 
currently powered by a hydrogen burning shell. 
\end{abstract}


\section{INTRODUCTION}

The study of globular cluster sdB stars has progressed
significantly in the last few years from both a observational and
theoretical perspective (see the extensive Introduction
in \citet{MB2006}). A particularly important question
whether short-period, close binaries are very common \citep{max01} 
or very rare \citep{MB2006} among the sdB stars 
urgently requires resolution; note that the first paper deals 
with the field objects while the latter is related 
to the population of the globular cluster NGC~6752.
It is not excluded that environmental effects are responsible 
for the observed difference. Based on the results from 
the SPY radial velocity survey, \citet{napiwotzki04} obtained 
39\% for the relative frequency of spectroscopic sdB
binaries among field stars. 

As part of the Las Campanas CASE project, a study of eclipsing binaries
in globular clusters \citep{kal05}, we present in this
paper photometric and spectroscopic observations of the eclipsing
sdB binary V209 in the globular cluster $\omega$~Cen. 
The binary V209~$\omega$~Cen in the catalogue of  
\citet{clement01} (also known as OGLE--GC14; from now on
called V209) was discovered
by \citet{kal96} during a survey for variable stars 
in the field of the globular cluster $\omega$~Cen. 
They presented a $V$-band light curve of the variable
and found an orbital period of $P=0.8344$~d. 
Further $BV$ photometry of V209 together with a  finding chart 
were published by \citet{kal04}.    
On the $V/(B-V)$ color-magnitude diagram of the cluster the variable
occupies a position at the top/blue edge of the blue straggler sequence.
In this paper we report photometric and spectroscopic
observations aimed at a determination of the absolute parameters of V209.

\section{PHOTOMETRIC OBSERVATIONS AND REDUCTIONS}

The photometric data were obtained with the 2.5-m du Pont telescope at
Las Campanas Observatory. A field of $8.65\times8.65$~arcmin was
observed with the TEK\#5 CCD camera at a scale of 0.259$\arcsec$/pixel.  
A large fraction of the frames were collected in a subraster mode with the
actual field of view reduced to $8.65\times4.32$~arcmin.
Observations were collected during the 1997, 1998 and 2003 
observing seasons. In 1997 and 1998 we used $BV$ filters 
while in 2003 we used mostly $VI$ filters with some additional $UB$ 
observations taken on a single night when 
several \citet{landolt92} photometric standard fields were also observed.
A summary of the photometric data is  
given in Table~\ref{tab1}. 
The raw data were pre-processed with the IRAF-CCDPROC
package\footnote{IRAF is distributed by the National Optical Astronomy
Observatories, which are operated by the Association of Universities
for Research in Astronomy, Inc., under cooperative agreement with the
NSF.}.  Time series photometric data were
extracted using the ISIS-2.1 image subtraction package \citep{alard,alard2}. 
Our procedure closely followed that described
in detail by \citet{moch02}. Zero
points for the ISIS instrumental magnitude differential light curves 
were determined from the
template images using the DAOPHOT/ALLSTAR package \citep{stetson}.
Aperture corrections were measured with the DAOGROW program \citep{stet90}. 
When determining the photometric  zero points we took
into account the flux from a close visual companion to V209.
This companion is located about 0.1 arsec East and 0.5 arcsec South
of the variable and has $V=20.77\pm 0.10$, $B-V=0.53\pm 0.27$, and 
$V-I=0.66\pm 0.16$. The companion was unambiguously detected on 
the stacked template images with  seeing of 0.67--0.70 arcsec. 
To minimize the effects of a variable point spread function we used
$600\times 600$ pixel sub-images in the ISIS analysis. In order  
to reduce the effects of crowding, the variable
was located in the south-west corner of the analyzed field;  
at the position of V209 in the cluster 
there is still a noticeable gradient of the stellar surface density. The  
lower crowding helped in the measurement of accurate aperture corrections.
\subsection{Photometric Calibration and the Light Curve}
\begin{figure}
\figurenum{1}
\label{fig1}
\plotone{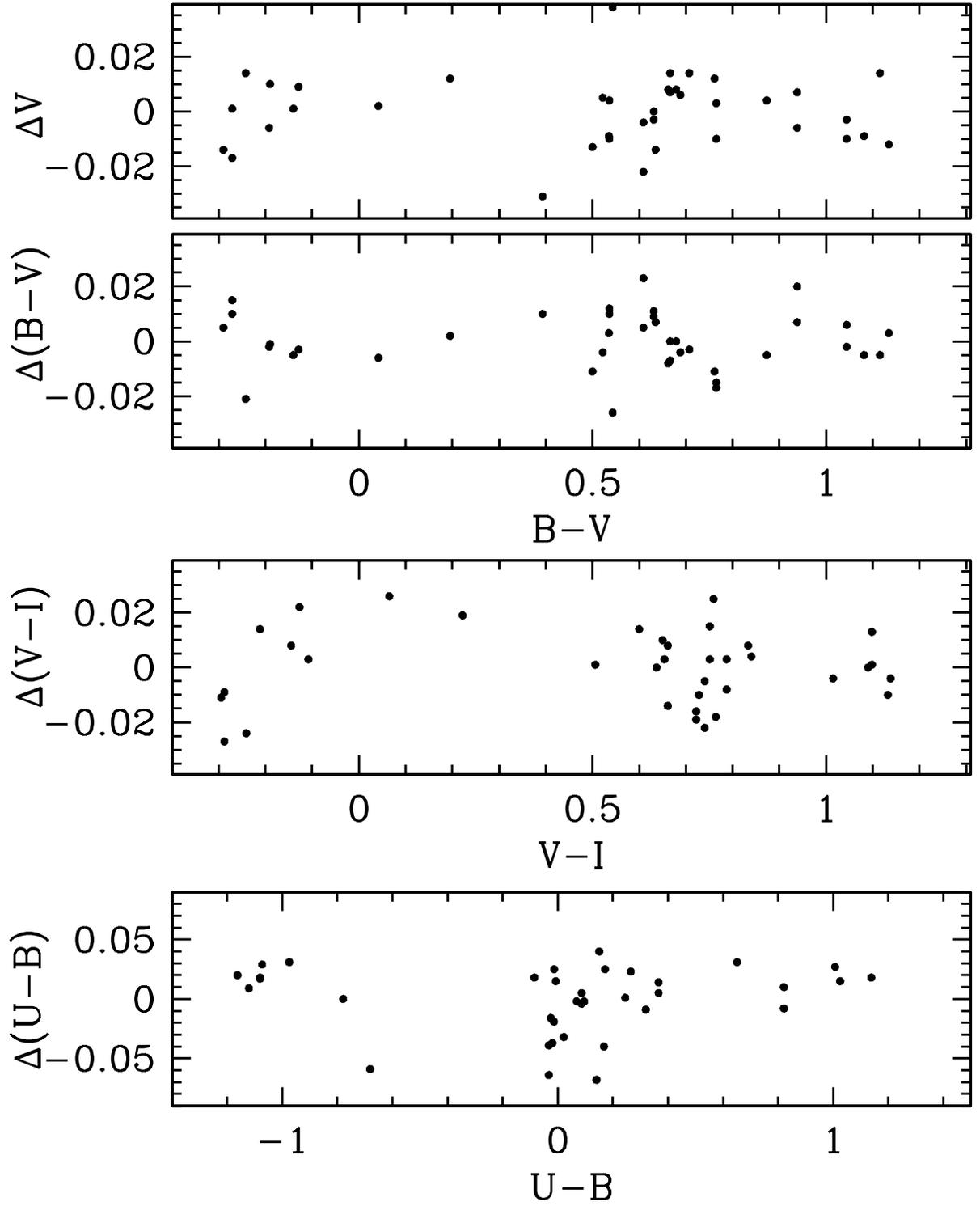}
\caption{
A plot of the color and magnitude residuals for the
Landolt standard stars observed on the night of 2003 May 4.
}
\end{figure}

We observed the field of V209 along with several \citet{landolt92}
fields on the photometric night 2003 May 4/5.
Specifically, 9 $UBVI$ observations
of 7 standard fields were collected, with two fields observed twice.
In total, we obtained 39 $UBVI$ measurements 
for 31 standard stars.
The standards were observed with a  range of air-mass $1.06<X<1.80$
while the variable was observed at $X=1.06$.
In Figure~\ref{fig1} we show the residuals between the standard and recovered
magnitudes and colors for the observed Landolt standards.
The total uncertainties of the zero points of our photometry 
are 0.02~mag for $BVI$ magnitudes and 0.05 mag for the $U-B$ color.
This estimate has been verified by a comparison
with  independently calibrated $BV$ photometry for the same
field collected during the 1998 season.
The differences of the zero points between these two sets, 
calculated for 190 stars with $14<V<18$ 
amount to $\delta V=0.000 \pm 0.015$ and $\delta(B-V)=0.014 \pm 0.013$.

\begin{figure}
\figurenum{2}
\label{fig2}
\plotone{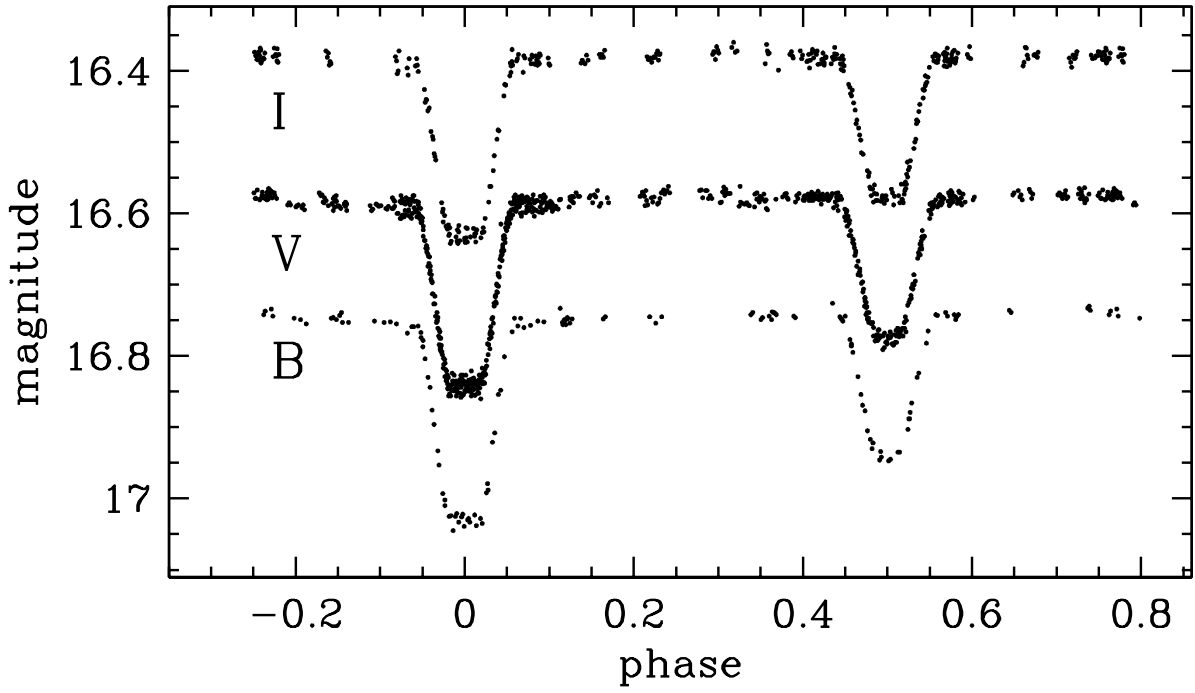}
\caption{
Phased $BVI$ light curves of V209.
}
\end{figure}

\begin{figure}
\figurenum{3}
\label{fig3}
\plotone{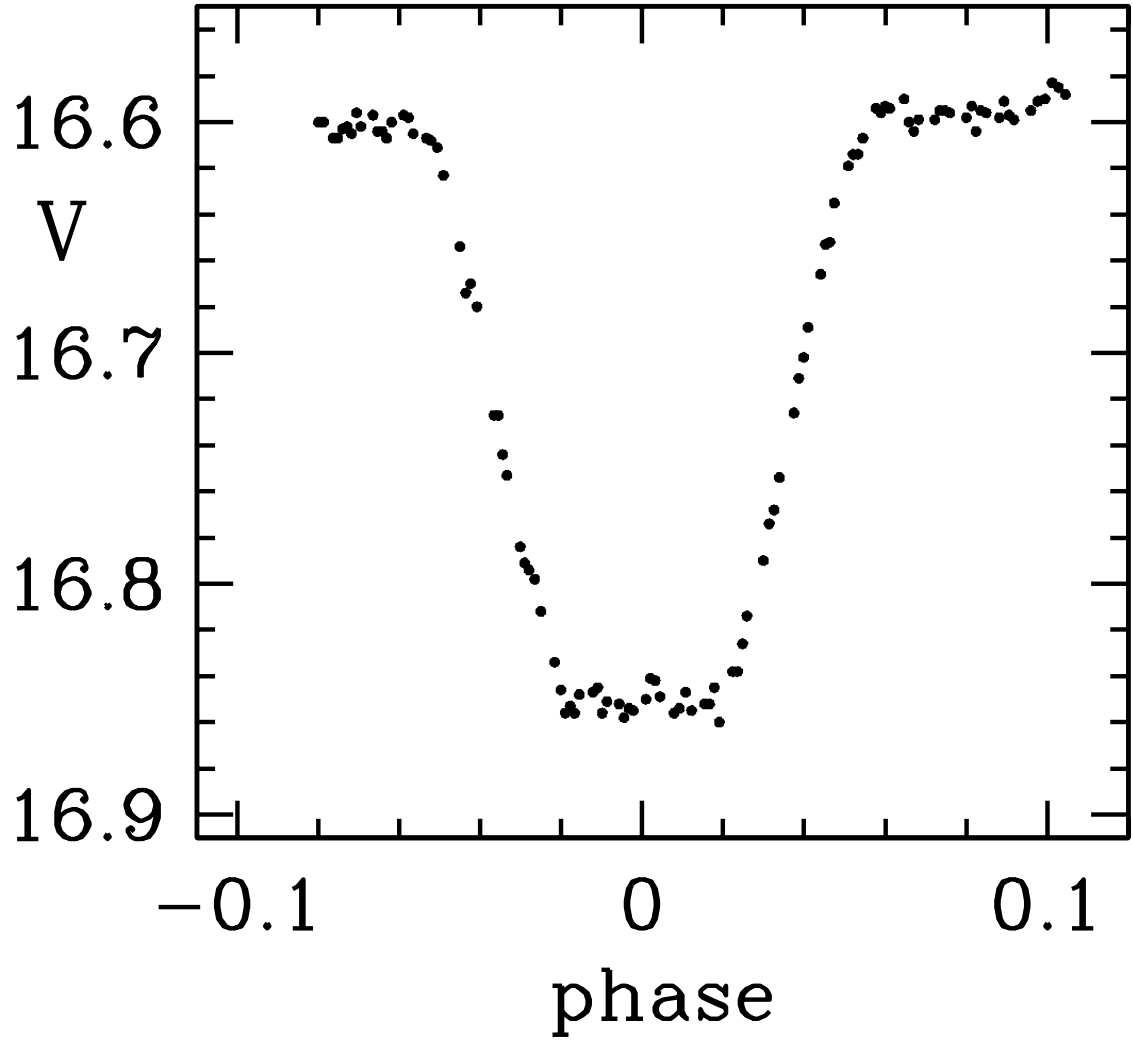}
\caption{
The $V$-band light curve of V209 in the primary eclipse.
}
\end{figure}

In Figure~\ref{fig2} we show $BVI$ 
light curves of V209 phased with the ephemeris
given in the following subsection. The variable exhibits 
only modest changes of $(B-V)$ and $(V-I)$ over the orbital cycle, 
with the colors becoming slightly
redder during the primary eclipse, as is normally observed
for distorted and gravity darkened components. 
The colors and magnitudes of V209
at  minima and at quadratures are listed in Table~\ref{tab2}. 
In addition, 
we have measured $U-B=-0.023\pm 0.009 $ at phase 0.36. 
The primary eclipse is total with the phase of constant light 
lasting about $0.040 P$.
This can be seen in Figure~\ref{fig3} which shows a 
particularly well covered eclipse event. 
The small out-of-eclipse curvature of the light curve indicates 
that to first order proximity 
effects due to ellipsoidal shape of the components 
and the reflection effect are negligible. Therefore we adopt the
magnitudes and colors observed at phase 0.0 as corresponding to the
larger component of the system (with the convention that the 
primary component is the one with larger size and  luminosity). 

The light curve solution presented in Sec.~4 implies that the
reflection effect in fact slightly modulates the light contributions 
of the individual binary components over the orbital period. 
Therefore, we did not attempt to measure the magnitudes and 
colors of the secondary component by subtracting the flux of the 
primary from the combined flux at maximum light.

\subsection{Period Study}

The observations collected with the du Pont telescope presented in this paper
together with other data described in \citet{kal96}
as well as some unpublished observations collected with the 1-m Swope 
telescope at Las Campanas Observatory cover 9 individual primary eclipses.
The times for these eclipses along with the timing
errors determined 
using the method of \citet{kwee} are given in Table~\ref{tab3}.

The O-C values listed in Table~\ref{tab3} correspond to the linear ephemeris:
\begin{equation}
Min I = HJD~245 0963.67780(5) + 0.83441907(4) 
\end{equation}
determined from a least squares fit to the time-of-eclipse data. A linear
ephemeris provides a good fit and there is no evidence for any
detectable period change during the interval 1993--2003 covered by the
data.

\section{SPECTROSCOPIC OBSERVATIONS AND REDUCTIONS}

Spectroscopic observations of V209  were obtained
with the MIKE echelle spectrograph 
\citep{bern03} on the Magellan~II
(Clay) telescope at Las Campanas Observatory.
The data were collected during observing runs in March and June of 2004.

For this analysis we only use data obtained with the blue channel of MIKE,
covering a range of 380 to 500 nm
at a resolving power of $\lambda / \Delta \lambda \approx 38,000$.
All of the observations were obtained with a $0.7\times 5.0$ arcsec slit
and with $2\times 2$ pixel binning. At 4380 \AA\ the resolution was
$\sim$2.7 pixels at a scale of 0.043~\AA/pixel.
The seeing ranged from 0.7 to 1.3~arcsec.
The spectra were first processed using a pipeline developed by Dan
Kelson following the formalism of \citet{kel03,kel06} 
and then analyzed further using standard tasks 
in the IRAF/Echelle package. Each of
the final  spectra consisted of two 1200~s exposures
interlaced with an exposure of a thorium-argon lamp. We obtained 11
spectra of V209. 
For the wavelength interval  400--500 nm, the average signal-to-noise 
ratio ranges from 25 to 45 depending on the observing conditions.
In addition to observations of V209, we also
obtained high S/N  spectra of HD~4850 to be used as radial velocity templates.
\subsection{Spectroscopic orbit of V209}
\begin{figure}
\figurenum{4}
\label{fig4}
\plotone{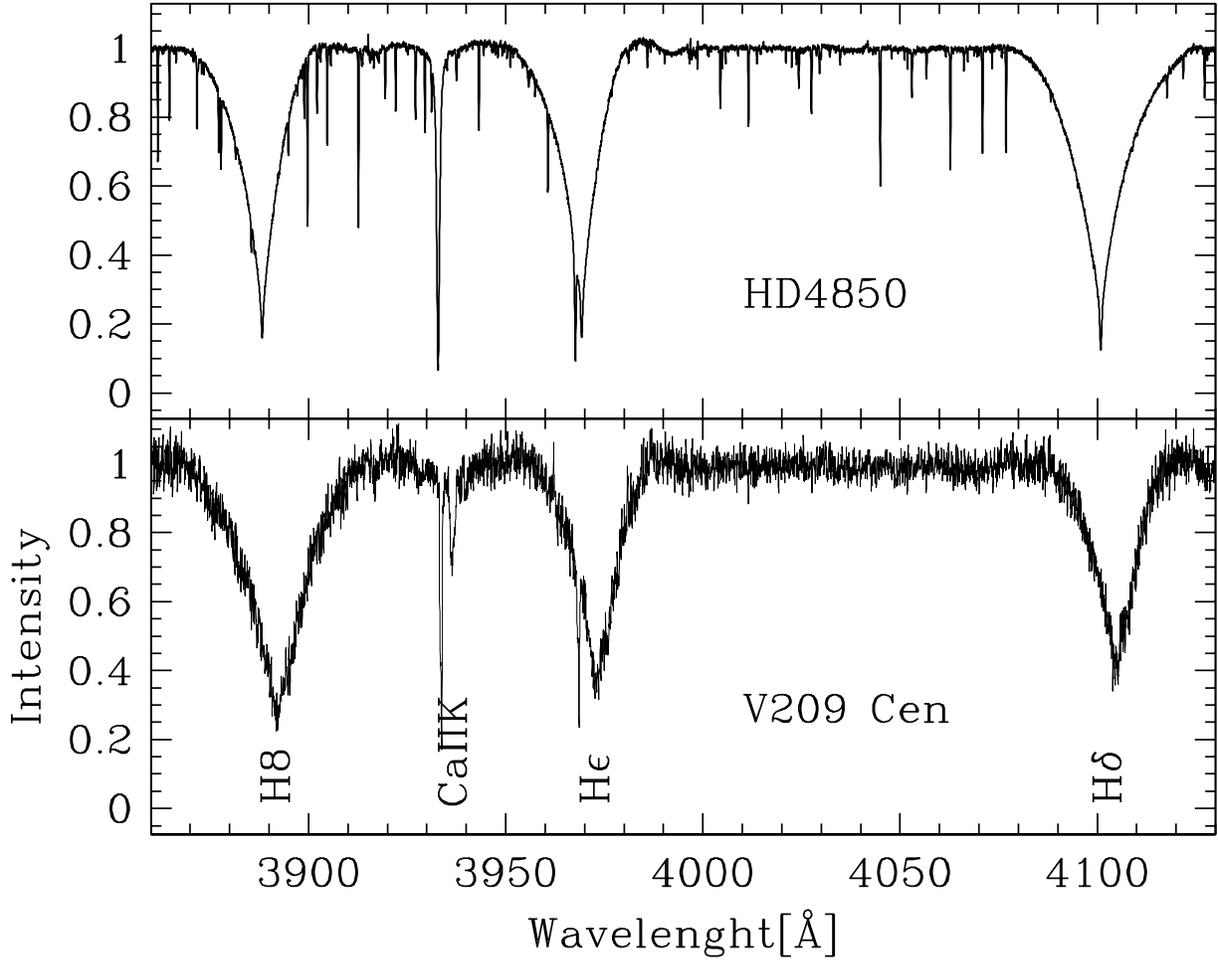}
\caption{
A sample spectrum of V209 taken at phase 0.755 (bottom) 
along with a spectrum of the template star (top).
}
\end{figure}

In Figure~\ref{fig4} we show a section of a spectrum of V209 obtained near  
the orbital quadrature at phase 0.755 along with the spectrum of the 
template star. The only easily identifiable absorption feature
besides the Balmer series lines is the Ca~II~K line located next to the 
interstellar line of the same ion; the narrow line inside H$\epsilon$ profile
is due to the interstellar Ca~II~H line. 
The hydrogen lines are broad and the components from the
individual component stars are strongly blended.
We have analyzed the spectra of V209 using 
the broadening function (BF) formalism \citep{ruc02}.
The BF analysis lets us study the effects of the spectral
line broadening and binary revolution splitting even for relatively
complicated profiles of the spectral lines. In the particular
case of V209, the main advantage is through the use of
intrinsically broad lines of hydrogen which are normally
avoided in cross-correlation analyses.

\begin{figure}
\figurenum{5}
\label{fig5}
\plotone{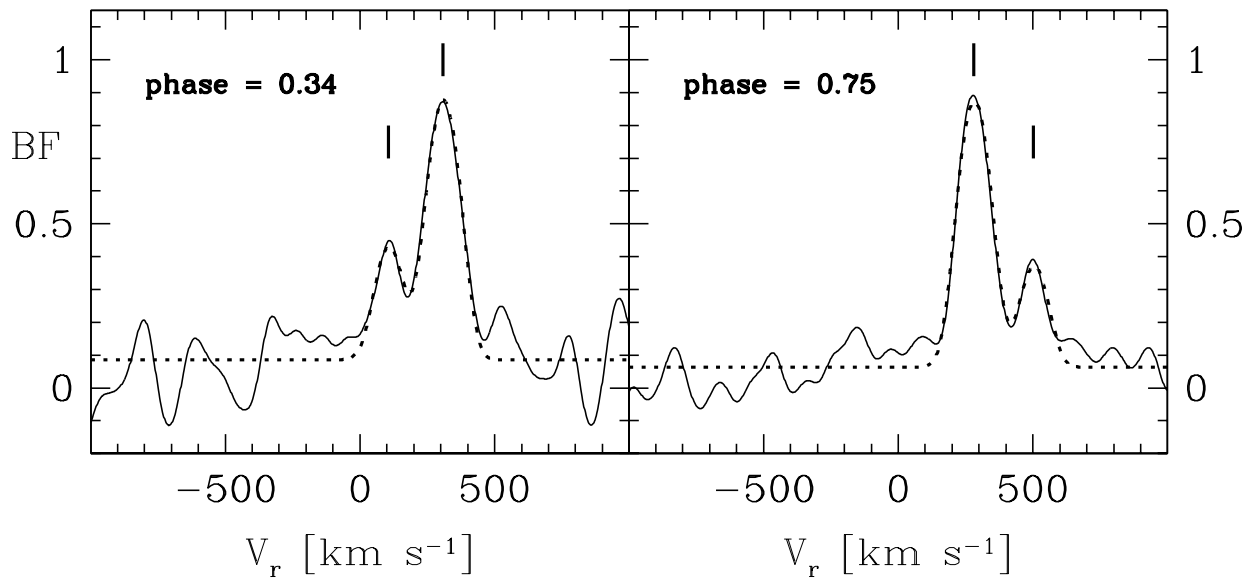}
\caption{
Broadening functions extracted from the spectra of V209 for obtained
at orbital phases of 0.34 (left panel) and 0.75 (right panel). The dashed 
line shows the fit of a model BF to the observed one. The vertical ticks
mark derived values of radial velocities. 
}
\end{figure}

\begin{figure}
\figurenum{6}
\label{fig6}
\plotone{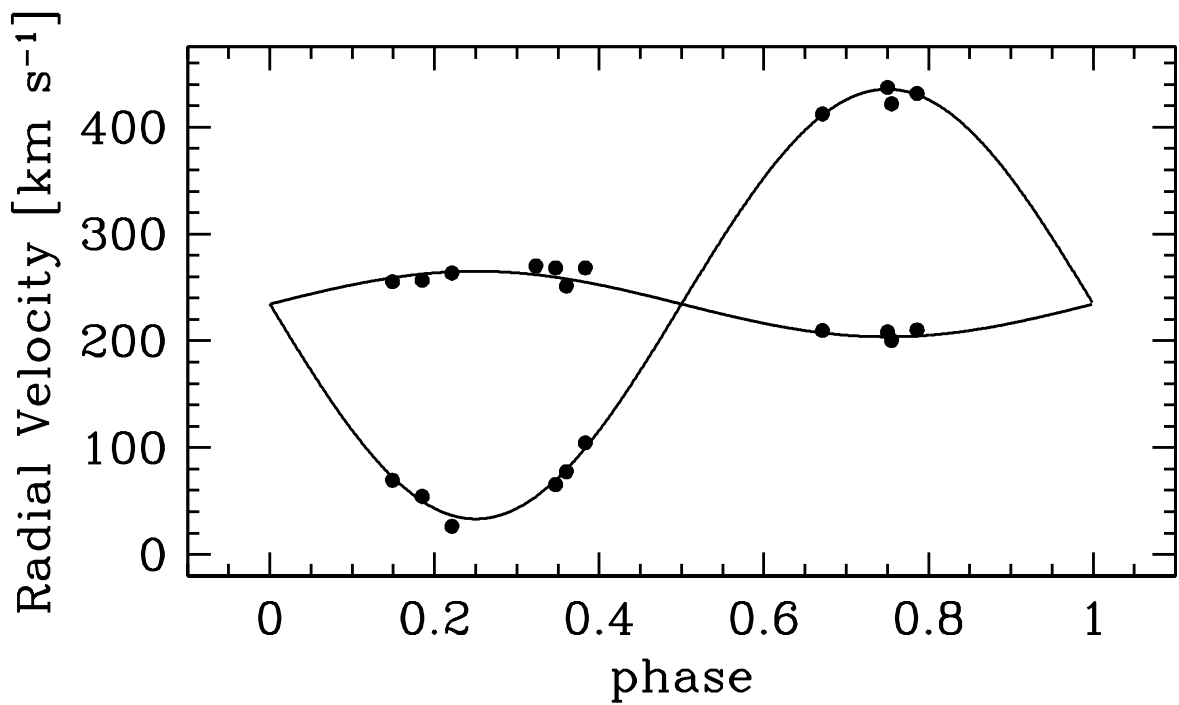}
\caption{
The spectroscopic observations and the adopted orbit for V209.
}
\end{figure}

A spectrum of HD~4850 ($V=9.64$, $B-V=0.03$, 
$Sp=A0V$ as given in SIMBAD)\footnote{This 
research has made use of the SIMBAD database,
operated at CDS, Strasbourg, France} was used as a BF method
template for V209.
The color of HD~4850 closely matches the unreddened color of V209;
we assumed a reddening of $E(B-V)$ = 0.131 \citep{schl98}.
According to \citet{kin00} the other relevant properties of 
HD~4850 are as follows: $V_{rad}=-41.7$~km/s, $V\sin i=14$~km/s, 
${\rm [Fe/H]}=-1.3$, $T_{eff}=8450$K and $\log (g)=3.20$.
We used the spectra in the wavelength range from 381 nm to 495 nm.
Figure~\ref{fig5} presents examples of fitting a model to 
the BFs calculated for two spectra taken near opposite quadratures;  
the procedure used to model the BF is discussed in some detail in 
\citet{kal06}. The synchronized velocities of rotation,
$V \sin i$ are 50 and 115 km~s$^{-1}$ for the primary
and secondary components, respectively, and these values are
confirmed by our BF modeling. 
Our velocity
measurements for V209 are listed in Table~\ref{tab4} where we give 
the heliocentric Julian date of mid-observation, the orbital 
phase, the measured radial velocity and the assigned weight for both
components and then 
the primary and secondary $(O-C)$ velocity deviations for the adopted orbit. 
The velocity observations and adopted orbit are presented in Figure~\ref{fig6}.
The current implementation of the BF method does not provide any
internal estimates of errors of the measured radial velocities. The weights
listed in Table~\ref{tab4} were assigned ``a posteriori'' based on the rms of 
the $O-C$ residuals for a given component and resulting from
the fitted spectroscopic orbit.
In practice we used an iterative approach starting from a 
spectroscopic solution with the assumed equal weights for both components. 
The velocity measured for the secondary components at phase 0.323 was 
assigned zero weight as it was derived from a relatively 
poorly defined peak in the BF.
 
A Keplerian orbit was fitted to the observations by fixing the period
and epoch to the precise ephemeris given above.
We assumed a circular orbit based on the photometric data.
The adjustable parameters in the orbital solution were the velocity
semi-amplitudes ($K_{1}$ and $K_{2}$) and the center of
mass velocity $V_0$. The fit was performed using the GAUSSFIT task within
IRAF/STSDAS.
The derived parameters of the spectroscopic orbit are listed in
Table~\ref{tab5}. The systemic velocity of the binary agrees with the radial
velocity of $\omega$~Cen, $v_{rad}=232.02$~km/s \citep{vandeven06}. 
At the location of V209 (about 7 arcmin 
from the cluster center) the velocity dispersion of $\omega$~Cen
is about 11.5~km/s (\citet{vandeven06}, Figure~\ref{fig3}).
We conclude that V209 is a radial velocity member of $\omega$~Cen.
\section{LIGHT CURVE SOLUTION}
We have analyzed the $BVI$ light curves of V209 using
the Wilson-Devinney model \citep{wd71} as implemented in the
light-curve analyzing program MINGA \citep{plewa}. The mass-ratio
of the binary was fixed at the spectroscopic value of $M_2/M_1 = 0.153
\pm 0.018$. The gravity darkening exponents and bolometric albedos  
were fixed at 1.0 and 1.0, respectively. The linear limb darkening
coefficients were adopted  from 
\citet{van93} for an assumed metallicity of ${\rm [Fe/H]=-1.7}$ 
which corresponds to the mean metallicity of the cluster 
\citep{stan06}.
The effective temperature of the primary component was estimated from
its color index, $B-V=0.183 \pm 0.020$, measured at the phase of  
totality in the primary eclipse.
The quoted uncertainty includes the external error arising from
the photometric calibration. \citet{schl98} extinction maps imply a 
reddening of $E(B-V)=0.131$, and using 
this value we obtain an unreddened 
color index for the primary component of $(B-V)_{0}=0.052\pm 0.020$. 
An estimate of effective temperature of $T_{1}=9370\pm 300$~K 
for the primary resulted from the use of the semi-empirical
calibration of \citet{vanden03}.

\begin{figure}
\figurenum{7}
\label{fig7}
\plotone{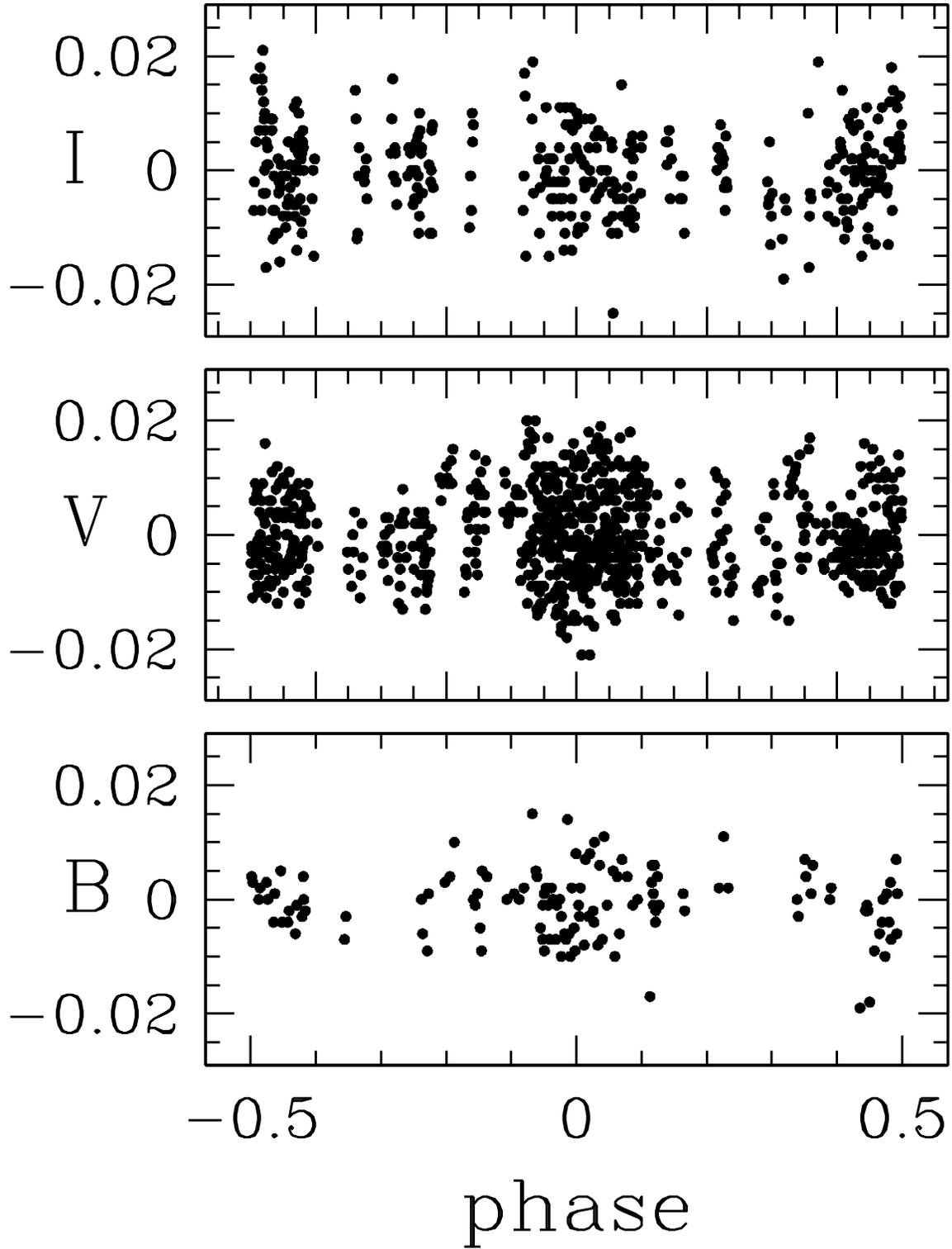}
\caption{
The residuals for the fit corresponding to the light curve
solution.
}
\end{figure}

The following parameters were adjusted in the light curve solution: 
the orbital inclination $i$, the non-dimensional potentials $\Omega_{1}$
and $\Omega_{2}$, the effective temperature of the secondary $T_{2}$,
and the relative luminosity of the primary $L_{1}(B;V;I)$. 
The  light curve solution obtained with MINGA is listed in 
Table~\ref{tab5}
and the residuals between the observed and synthetic light curves are 
shown in Figure~\ref{fig7}.
In Table~\ref{tab6} we give  the ``equal volume''
mean radii of the components. The  solution 
implies that   the difference between ``polar'' 
and the ``point'' radii amounts to about 2\% for both components, 
the two radii being the radius toward the stellar pole 
and toward the Lagrangian point L1 of the binary orbit.
The system has a detached configuration with both components showing
only a very modest ellipsoidal distortion.

\section{ABSOLUTE PARAMETERS}
The absolute parameters of V209 obtained from our spectroscopic and 
photometric analysis are given in Table~\ref{tab7}. The errors in the
temperatures include all sources of uncertainties. The absolute visual
magnitudes $M_{\rm V}$ were calculated by adopting bolometric corrections
$BC_{V1}=-0.22$ and $BC_{V2}=-0.52$ which were derived from relations 
presented by  \citet{vanden03}. 
There are two uncertainties involved in this step.
First, we assumed  $[{\rm Fe/H}]=-1.7$, representing the peak in 
the wide distribution of metallicities measured for the
$\omega$~Cen stars \citep{stan06}.
Second, the adopted bolometric corrections are appropriate for atmospheres
of ``normal'' main-sequence stars while the chemical composition of 
the components of V209 was almost certainly affected by the evolution of 
the binary through a common-envelope phase.
   
We estimate the observed visual magnitudes
of  the components of V209 from the light curve solutions
at maximum light as
$V_1=16.825\pm 0.021$, $B_1=17.015\pm 0.021$, and
$V_2=18.317\pm 0.022$, $B_2=18.382\pm 0.029$,  where the
uncertainties include the errors of the photometric zero point.
This leads to apparent distance moduli $(m-M)_{\rm 1}=13.49\pm0.14$ 
and $(m-M)_{\rm 2}=13.51\pm0.12$ for components 1 and 2, respectively
assuming the reddening of $E(B-V)=0.131$ 
and using the absolute magnitudes from Table~\ref{tab7}.
Keeping in mind all of the sources of uncertainty mentioned above, 
we conclude 
that the derived distances are in reasonable agreement with other estimates 
of the cluster distance. For example \citet{vandeven06}, using 
dynamical data, recently measured  $(m-M)_{0}=13.75\pm 0.13$.
Our estimate of the distance of V209 provides an additional evidence that
the binary belongs to $\omega$~Cen.
Figure~\ref{fig8} shows the location of the individual components of
the binary on a color-magnitude diagram (CMD) of the cluster.
The less massive and hotter component of V209 is located 
on the CMD in the area occupied by stars from 
the extreme horizontal branch of the cluster. Its more massive companion
is located between the  blue stragglers region and the blue horizontal 
branch.  
\section{DISCUSSION AND SUMMARY}
In order to discuss the evolutionary status of V209 we compare
the location of its components in the $T_{eff}-\log(g)$ diagram  
to that of several sdB stars in close binary systems in Figure~\ref{fig9}. 
Data for the sdB stars were taken from \citet{max01} and
\citet{morales03}. 
A sample of blue horizontal branch stars in globular 
clusters was taken from \citet{behr2003}. The individual objects
marked in Figure~\ref{fig9} include the visible component of the 
binary stars HD~188112 \citep{heber03}, the 
helium white dwarf SDSS~J123410.37--022802.9
\citep{liebert04}, the visible component of the close binary 
M4--V46(NGC~6121--V46) \citep{o'toole06} and the white dwarf SDSS~J0917+46
\citep{kilic06}. These four objects have estimated masses
spanning the range $0.17-0.24\,M_{\odot}$. 
Also shown are evolutionary tracks for low-mass, post-red giant 
branch stars which lost their envelopes before core helium ignition 
\citep{driebe98} as well as a location of zero-age (ZAHEB) and terminal-age 
(TAEHB) extreme horizontal branches \citep{dorman93}.  
The components of V209 have lower densities and
effective temperatures  compared with sdB stars in binaries.
The secondary component of V209 has an effective temperature and a 
bolometric luminosity both too low for the star to be classified as a normal 
extreme horizontal branch star. Its mass is low, but it is within the range
of masses observed for companions of white dwarfs in binaries 
\citep{nelemans05}. Hence, it may be considered as a low mass
star evolving toward the sdB or white dwarf stage. 

Two recently identified low-mass helium white dwarfs 
have masses comparable to the secondary of V209. 
\citet{o'toole06} estimated a mass of 0.17~$M_{\odot}$ for 
the field star SDSS~J0917+46. Nothing is know for the moment about 
possible binarity of this object. A photometric variable V46 in the 
globular cluster M4 was shown by \citet{o'toole06} to be
a single line spectroscopic binary whose visible component 
has a mass of about $0.19\,M_{\odot}$. The orbital period of this
system is about 2.1 hours and its invisible component is most likely
a white dwarf with a lower mass limit of only 0.26~$M_{\odot}$.

The mass of the primary component of V209 is too large
to be a recently formed post-AGB star from $\omega$~Cen.
The star is also more massive than most of helium white dwarfs
from the field population. The mass distribution obtained for
1175 DA white dwarfs from the SDSS sample by \citet{madej04}
shows a strong peak at 0.56~$m_{\odot}$  and contains only
about two dozen objects with masses exceeding 0.90~$m_{\odot}$.
The radius of the primary of V209 corresponds to that of a main sequence
star of its mass. However, its luminosity as well as  
effective temperature are far too high to consider it a normal dwarf.

It is widely accepted that close binaries with degenerate components
form through common-envelope (CE) evolution \citep{pacz76}. In
particular, \citet{han03} studied some specific channels  leading to
the formation of binaries with sdB stars. We propose that V209 was
formed through a process similar to the 
scenario called by these authors ``the second 
common-envelope ejection channel''. It applies to  systems  consisting
of a white dwarf (originally the more massive 
component of the binary) and a red giant. 
The binary evolved first through the CE phase when its primary
ascended the giant branch. As a result of angular momentum loss
the system became a close binary with an orbital period of the 
order of one day.  At this stage it reached a detached configuration and
consisted of a white dwarf and the  main sequence companion. The second
episode of the mass loss/transfer began once the companion entered the 
Hertzsprung gap. After losing most of its envelope it failed to 
ascend the red giant branch and to ignite helium in its core.
Instead, it started to evolve along the horizontal direction on
the H-R diagram moving toward the sdB domain. 
As for the primary component of the binary, it accreted an amount
of mass sufficient for formation and ignition of a hydrogen shell 
on its surface. This led  to an expansion of the photosphere and 
to the creation of a ``re-born'' star with an extended envelope. 
The former white dwarf is currently seen as the more massive and 
more luminous component of V~209 Cen while the
less evolved component has lost most 
of its envelope and is currently composed of a helium core surrounded 
by a thin hydrogen shell. It is now seen as the less massive and 
hotter component of the binary. This kind of evolution was modeled 
is some detail by \citet{burd02} and 
\citet{ergma03} who
presented a scenario explaining the observed properties of 
the optical companion of the millisecond pulsar PSR J1740-5340.

An accurate explanation of an evolutionary status 
of V209 deserves further attention but it is beyond the scope 
of this paper to provide a detailed model for this peculiar binary. 
Admittedly, the above proposed evolutionary scenario is a
highly speculative one. As it was pointed out by the referee,
our scenario implies that the current primary would 
evolve very rapidly across the H-R diagram so that
an {\it a priori\/} probability of seeing it at this
short lasting phase of evolution is small. 
On the other hand, we note that $\omega$~Cen a is an exceptionally
populous cluster with a rich population of hot subdwarfs. 
Also, we have failed to identify any other similar objects in the sample
of 12 clusters surveyed for photometric variables 
during the CASE project. Hence, it is feasible that V209 has been 
caught -- by a shear luck -- 
in this very short-lasting phase of its evolution.

\begin{figure}
\figurenum{8}
\label{fig8}
\plotone{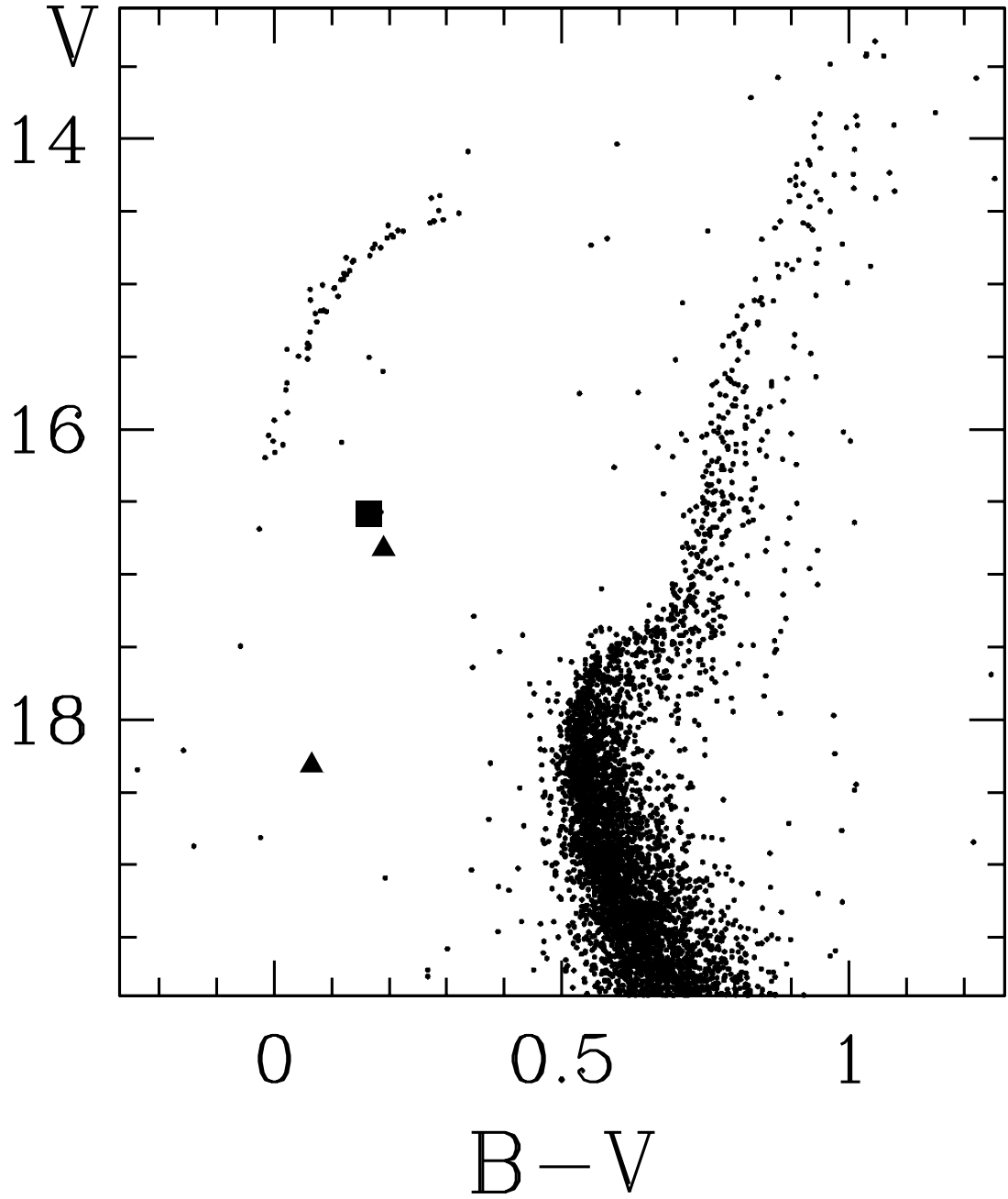}
\caption{
The color – magnitude diagram for $\omega$~Cen 
stars near V209. The locations 
of both components of the binary are marked 
(filled triangles) along with the 
position of the combined photometric data (filled square).}
\end{figure}

\begin{figure}
\figurenum{9}
\label{fig9}
\plotone{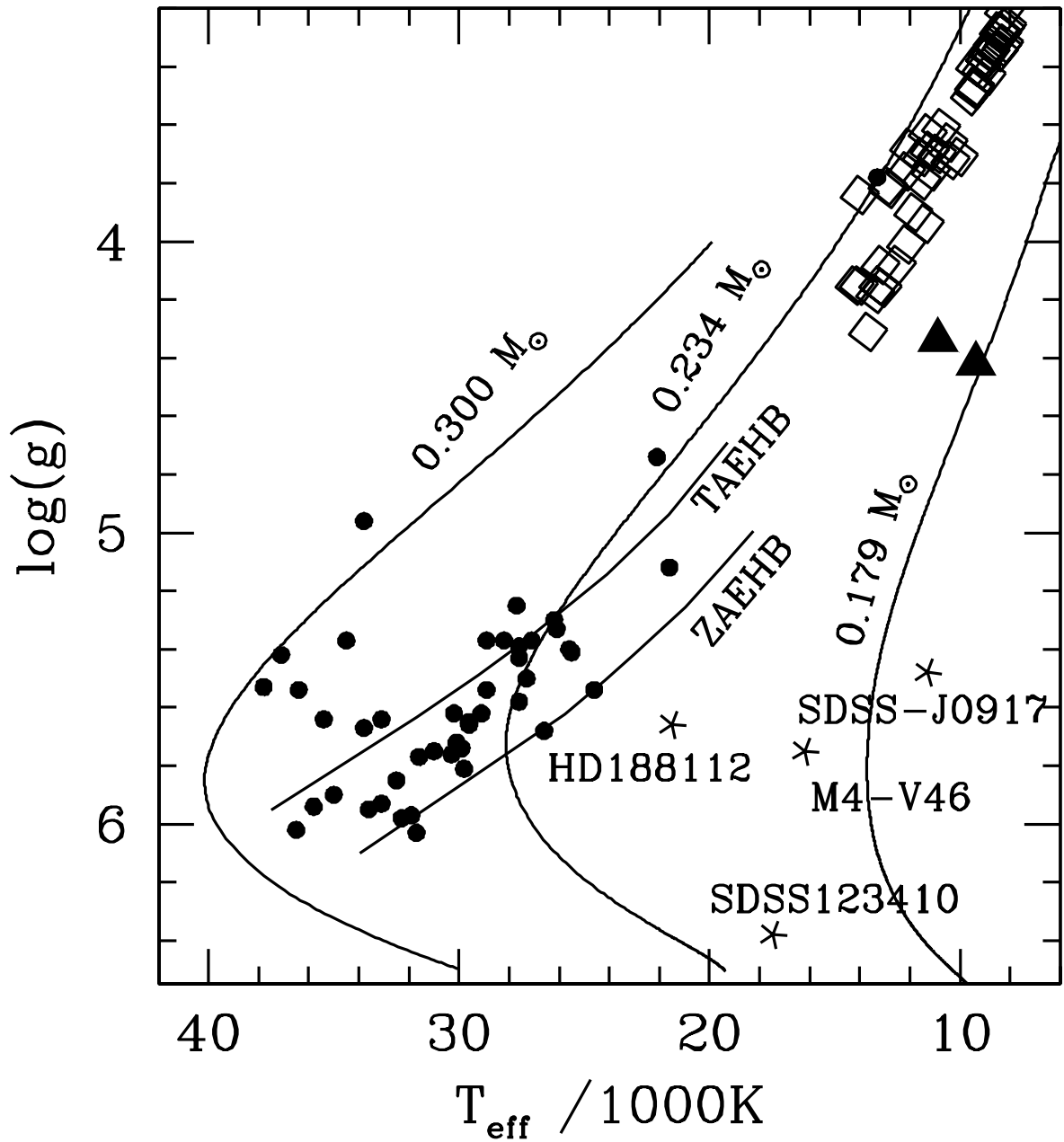}
\caption{The position of the components of V209 (triangles) in the 
($T_{eff}$, $\log g$) plane. The locations of some sdB stars 
in close binaries are marked with dots while the
open squares correspond to blue horizontal branch stars. 
The continuous lines 
labeled with stellar mass show evolutionary tracks for
low-mass post-red giant-branch stars. See the text for more details.
}
\end{figure}

In summary, V209~$\omega$~Cen, the second binary analyzed 
within the CASE project (after OGLEGC~17 = V212~$\omega$~Cen, 
\citet{tho2001,kal02}), appears to be a very unusual
star. In terms of its location on the color -- magnitude
diagram (Figure~\ref{fig8}) and its very highly
evolutionary modified properties,
it is a very different binary from the main targets 
of the CASE project which are Main Sequence, detached binaries
with unevolved and moderately evolved components. 
Such binaries, which do not experience the preferential
selection effects of V209, will be discussed in the
subsequent publications of the CASE project.

\acknowledgments
We thank Dr. U. Heber for sending us the data defining
location of model ZAEHB and TAEHB.

JK and WP were supported by the grants 1~P03D~001~28  and
76/E-60/SPB/MSN/P-03/DWM35/2005-2007 from the Ministry
of Science and Information Society Technologies, Poland. IBT was
supported by NSF grant AST-0507325.
Support from the Natural Sciences and Engineering Council of Canada
to SMR is acknowledged with gratitude.

\clearpage

\begin{deluxetable}{ccccc}

\tablecolumns{5}
\tablewidth{0pt}
\tabletypesize{\normalsize}
\tablecaption{Summary of Photometric Observations of V209
\label{tab1}}
\tablehead{
\colhead{Band}                  &
\colhead{N}                     &
\colhead{Exp Time}              &
\colhead{FWHM}                  &
\colhead{$<$FWHM$>$}            \\  
\colhead{}                      &
\colhead{}                      &
\colhead{sec}                   &
\colhead{arcsec}                &
\colhead{arcsec}
}
\startdata
U    & 6  & 100-180& 1.09-1.22&  1.14\\
B    & 140& 40-120 & 0.79-1.71&  1.11\\
V    & 760& 20-100 & 0.66-1.76&  1.09\\
I    & 355& 17-70  & 0.67-1.39&  0.98\\

\enddata

\end{deluxetable}


\begin{deluxetable}{lccccc}
\tablecolumns{6}
\tablewidth{0pt}
\tabletypesize{\normalsize}
\tablecaption{$BVI$ Photometry of V209 at Minima and Quadrature
\label{tab2}}
\tablehead{
\colhead{Phase}            &
\colhead{$V$}              &
\colhead{$B$}              &
\colhead{$I$}              &
\colhead{$B-V$}            &  
\colhead{$V-I$}                      
}
\startdata
Max & 16.580 & 16.744 & 16.380 & 0.164 & 0.200 \\
Min~I & 16.846 & 17.029 & 16.632 & 0.183 & 0.214 \\
Min~II & 16.779 & 16.943 & 16.578 & 0.164 & 0.201 \\
\enddata

\end{deluxetable}

\begin{deluxetable}{rccc}
\tablecolumns{4}
\tablewidth{0pt}
\tabletypesize{\normalsize}
\tablecaption{Times of Minima and $O-C$ Values for V209
\label{tab3}}
\tablehead{
\colhead{Cycle}            &
\colhead{$T_0$}            &
\colhead{Error}            &
\colhead{$O-C$}            \\  
\colhead{}                 &  
\colhead{HJD-2400000}      &
\colhead{}                 & 
\colhead{}                     
}
\startdata
-2255.5 & 49081.6456 & 0.0003 &  0.0000\\
 -434.0 & 50601.5404 & 0.0002 & -0.0005\\
    0.0 & 50963.6777 & 0.0001 &  0.0001\\
    3.5 & 50966.5983 & 0.0001 &  0.0000\\
    6.0 & 50968.6842 & 0.0001 &  0.0001\\
 2156.0 & 52762.6859 & 0.0005 & -0.0006\\
 2158.5 & 52764.7713 & 0.0002 &  0.0001\\
 2159.5 & 52765.6057 & 0.0002 &  0.0001\\
 2163.0 & 52768.5263 & 0.0001 &  0.0000\\
\enddata

\end{deluxetable}

\eject

\begin{deluxetable}{ccccccrr}
\tablecolumns{8}
\tablewidth{0pt}
\tabletypesize{\normalsize}
\tablecaption{Radial Velocities of V209 and Residuals 
from the Adopted Spectroscopic Orbit \label{tab4}}
\tablehead{
\colhead{HJD-2450000}       &
\colhead{phase}             &
\colhead{RV$_1$}            &
\colhead{w$_1$}             &  
\colhead{RV$_2$}            &  
\colhead{w$_2$}             &
\colhead{$(O-C)_1$}         & 
\colhead{$(O-C)_2$}                     
}
\startdata
3182.5224 & 0.149 & 255.51& 0.85& 69.43 &1.0 &-3.62 & -3.04\\
3182.5528 & 0.185 & 256.66& 0.85& 54.37 &1.0 &-5.92 & 4.51\\
3182.5829 & 0.221 & 263.42& 0.85& 26.31 &1.0 &-1.17 & -10.34\\
3178.4959 & 0.323 & 270.18& 0.85& 99.67 &0.0 &8.25  & 45.56\\
3066.7032 & 0.347 & 268.31& 0.85& 65.56 &1.0 &8.73  & -3.93\\
3178.5260 & 0.360 & 251.13& 0.85& 77.55 &1.0 &-6.76 & -1.92\\
3066.7334 & 0.383 & 268.33& 0.85&104.60 &1.0 &13.33 & 5.05\\
3179.6203 & 0.671 & 209.70& 0.85&412.33 &1.0 &2.24  & 1.06\\
3068.7084 & 0.750 & 208.40& 0.85&437.29 &1.0 &4.64  & 1.75\\
3180.5251 & 0.755 & 200.47& 0.85&421.91 &1.0 &-3.31 & -13.53\\
3068.7386 & 0.786 & 210.14& 0.85&431.76 &1.0 &5.60  & 1.18\\
\enddata

\end{deluxetable}

\begin{deluxetable}{lr}
\tablecolumns{2}
\tablewidth{0pt}
\tabletypesize{\normalsize}
\tablecaption{Orbital Parameters for V209
\label{tab5}}
\tablehead{
\colhead{Parameter}       &
\colhead{Value}             
}

\startdata
$P$ (days) & 0.83441907(fixed) \\ 
$T_{0}$~(HJD-245 0000) & 963.67780(fixed) \\
$V_0~(km~s^{-1})$ & 234.43$\pm$ 1.49 \\
$e$    & 0.0(fixed) \\
$K_{1}~(km~s^{-1})$  & 30.67$\pm$2.39 \\
$K_{2}~(km~s^{-1})$  & 201.11$\pm$2.31 \\
Derived quantities: & \\
$A \sin i$~($R_{\odot}$) & 3.824$\pm$0.055\\
$M_{1} \sin^3 i$~($M_{\odot}$) & 0.934$\pm$0.040 \\
$M_{2} \sin^3 i$~($M_{\odot}$) & 0.142$\pm$0.006 \\
\enddata

\end{deluxetable}

\begin{deluxetable}{lr}
\tablecolumns{2}
\tablewidth{0pt}
\tabletypesize{\normalsize}
\tablecaption{Light Curve Solution for V209
\label{tab6}}
\tablehead{
\colhead{Parameter}       &
\colhead{Value}             
}

\startdata
$i$~(deg) & 85.03 $\pm$ 0.21 \\
$\Omega_{1}$ & 4.091 $\pm$ 0.025 \\
$\Omega_{2}$ & 2.815 $\pm$ 0.017 \\
$T_{1}$~(K) & 9370 (fixed) \\
$q=M_2/M_1$ & 0.153 (fixed) \\
$T_{2}$~(K) & 10866 $\pm$ 122 \\
$(L_{1B}/L_{2B})$ &3.52  $\pm$ 0.09 \\
$(L_{1V}/L_{2V})$ &3.95  $\pm$ 0.10 \\
$(L_{1I}/L_{2I})$ &4.03  $\pm$ 0.19 \\
$<r_{1}>$ & 0.2560 $\pm$ 0.0016\\
$<r_{2}>$ & 0.1106 $\pm$ 0.0013\\
rms~(B)~(mag) & 0.0059 \\
rms~(V)~(mag) & 0.0075 \\
rms~(I)~(mag) & 0.0074 \\
\enddata

\end{deluxetable}

\begin{deluxetable}{lr}
\tablecolumns{2}
\tablewidth{0pt}
\tabletypesize{\normalsize}
\tablecaption{Absolute Parameters for V209
\label{tab7}}
\tablehead{
\colhead{Parameter}       &
\colhead{Value}             
}

\startdata
$A$~($R_{\odot}$)      & 3.838$\pm$0.055\\
$M_{1}$~($M_{\odot}$)  & 0.945$\pm$0.043 \\
$M_{2}$~($M_{\odot}$)  & 0.144$\pm$0.008 \\
$R_{1}$~($R_{\odot}$)  & 0.983$\pm$0.015 \\
$R_{2}$~($R_{\odot}$)  & 0.425$\pm$0.008 \\
$T_{1}$~(K)            & 9370$\pm$300 \\
$T_{2}$~(K)            & 10866$\pm$323 \\
$L^{bol}_{1}$($L_{\odot}$)&6.68 $\pm$0.88\\
$L^{bol}_{2}$($L_{\odot}$ &2.26 $\pm$0.28\\
$M_{\rm V1}~(mag)$     &2.90 $\pm$0.13\\
$M_{\rm V2}~(mag)$     &4.37 $\pm$0.12\\
${\rm log}~g_{1}({\rm cm~s^{-2}})$&4.43 $\pm$0.02\\
${\rm log}~g_{2}({\rm cm~s^{-2}})$&4.34 $\pm$0.02\\
\enddata

\end{deluxetable}

\clearpage
 
\end{document}